\documentclass[aps,prl,showpacs,twocolumn,superscriptaddress]{revtex4}
\usepackage{amsmath}
\usepackage{amssymb}
\usepackage{epsfig}
\usepackage{color}
\usepackage{graphicx,amsmath}

\begin{document}
\title{Quasi-one-dimensional quantum
spin liquid in the $\rm {\bf Cu(C_4H_4N_2)(NO_3)_2}$ insulator}
\author{V. R. Shaginyan}\email{vrshag@thd.pnpi.spb.ru}
\affiliation{Petersburg Nuclear Physics Institute, NRC
Kurchatov Institute, Gatchina, 188300,
Russia}\affiliation{Clark Atlanta University, Atlanta, GA
30314, USA}
\author{V. A. Stephanovich}\email{stef@math.uni.opole.pl}
\affiliation{Institute of
Physics, Opole University, Opole, 45-052, Poland}
\author{K. G. Popov}\affiliation{Komi Science
Center, Ural Division, RAS, Syktyvkar, 167982, Russia}
\author{E. V. Kirichenko}\affiliation{Institute of Mathematics
and Informatics, Opole University, Opole, 45-052, Poland}

\begin{abstract}

We analyze measurements of the magnetization, differential
susceptibility and specific heat of quasi-one dimensional
insulator Cu(C$_4$H$_4$N$_2$)(NO$_3$)$_2$ (CuPzN) subjected to
magnetic fields. We show that the thermodynamic properties are
defined by quantum spin liquid formed with spinons, with the
magnetic field tuning the insulator CuPzN towards quantum
critical point related to fermion condensation quantum phase
transition (FCQPT) at which the spinon effective mass diverges
kinematically. We show that the FCQPT concept permits to reveal
and explain the scaling behavior of thermodynamic
characteristics. For the first time, we construct the schematic
$T-H$ (temperature---magnetic field) phase diagram of CuPzN,
that contains Landau-Fermi-liquid, crossover and non-Fermi
liquid parts, thus resembling that of heavy-fermion compounds.
\end{abstract}


\maketitle

Recently, the striking measurements of the thermodynamic
properties at low temperatures $T$ under the application of
magnetic field $H$ on the quasi-one dimensional (Q1D) insulator
Cu(C$_4$H$_4$N$_2$)(NO$_3$)$_2$ (CuPzN) have been performed
\cite{prl15}. The observed thermodynamic properties of CuPzN is
very unusual and nobody expects that it might belong to the
class of HF compounds, including quasicrystals (QC)
\cite{shext}, insulators with quantum spin liquid (QSL), and
heavy-fermion (HF) metals \cite{quasi,pr,herb2011,book}. Similar
Q1D clean HF metal $\rm YbNi_4P_2$ was recently experimentally
studied, that reveals it has a Q1D electronic structure and
strong correlation effects dominating the low-temperature
properties, while its thermodynamic properties resembles those
of HF metals, including the formation of Landau-Fermi-liquid
(LFL) ground state \cite{steg}. These observations show that
both CuPzN and $\rm YbNi_4P_2$ can demonstrate a new type of Q1D
Fermi liquid whose thermodynamic properties resemble that of HF
compounds rather than the Tomonaga-Luttinger system. One of the
hallmark features of geometrically frustrated insulators is
spin-charge separation. The behavior of Q1D Fermi liquid (Q1DFL)
is the subject of ongoing intensive experimental research in
condensed matter physics, see, e.g. \cite{prl15} and references
therein. Q1DFL survives up to the saturation magnetic field
$H_s$, where the quantum critical point (QCP) occurs giving way
to a gapped, field-induced paramagnetic phase \cite{prl15}. In
other words, at $H=H_s$ both antiferromagnetic (AFM) sublattices
align in the field direction i.e. the magnetic field fully
polarizes Q1DFL spins. We will see below that in Q1DFL the
fermion condensation quantum phase transition (FCQPT) plays a
role of QCP, at which the energy band for spinons becomes almost
flat at $H=H_s$ and the effective mass $M^*$ of spinons diverges
due to kinematic mechanism. Thus, the bare interaction of
spinons is weak \cite{prl15}. In that case the original
Tomonaga-Luttinger system can exactly be mapped on a system of
free spinons, which low-temperature behavior in magnetic fields
can be viewed as the LFL one \cite{rojkov}. Thus CuPzN offers a
unique possibility to observe a new type of Q1D QSL whose
thermodynamic properties resemble that of HF compounds like HF
metals, including Q1D HF metal $\rm YbNi_4P_2$ \cite{steg},
quantum spin liquids of herbertsmithite ZnCu$_3$(OH)$_6$Cl$_2$
\cite{herb2011}, and liquid $^3$He \cite{prl08}. Theory of Q1D
liquids is still under construction and recent results show that
the liquids can exhibit LFL, non-Fermi liquid (NFL) and
crossover behavior \cite{rojkov,rozhkov14,lebed}.

In this letter we show that, contrary to ordinary wisdom, CuPzN
can be regarded as an insulator belonging to HF compounds, while
its thermodynamic properties are defined by weakly interacting
Q1D QSL formed with spinons, and are similar to those of HF
compounds. Here spinons are chargeless fermionic quasiparticles
with spin $1/2$. For the first time, we demonstrate that its
$T$-$H$ phase diagram contains LFL, crossover and NFL parts,
thus resembling that of HF compounds. To unveil the relation
between CuPzN and HF compounds, we study the scaling behavior of
its thermodynamic properties which are independent of the
interparticle interaction. We demonstrate that CuPzN exhibits
the universal scaling behavior, that is typical of HF compounds.

Upon transition to fermionic description, CuPzN is indeed
represented by weakly interacting fermions. The description of
weakly interacting fermion gas gives magnetization in terms of
fermion number per spin $N/L=\int_0^\infty
D(\varepsilon)f(\varepsilon-\mu(H)) d\varepsilon$, where $L$ is
the number of spins in Q1D chain, $D(\varepsilon)$ is the
density of states, corresponding to free fermion spectrum
$\varepsilon=p^2/(2m_0)$ with $p$ is the momentum and $m_0$ is
the bare mass. The chemical potential $\mu(H)=H_s-H$, and
$f(x)=(e^x+1)^{-1}$ is the Fermi distribution function
\cite{prl15,prl07,prl00}. The magnetization can be expressed as
$M=M_s-N$ ($M_s$ is the saturation magnetization) or explicitly
\begin{equation}\label{tll2}
M(H,T)=M_s-\frac{\sqrt{2m_0T}}{\pi}\int_0^\infty
\frac{dx}{e^{\left(x^2-\frac{H_s-H}{T}\right)}+1}.
\end{equation}
Equation \eqref{tll2} will be used below to calculate the
differential magnetic susceptibility $\chi(T,H)=\frac{\partial
M(T,H)}{\partial H}$. To calculate the specific heat $C(T,H)$,
we need the internal energy $E$, which in the above approach can
be calculated as follows
\begin{equation}\label{ET}
E(T,H)=\int_0^\infty \varepsilon
D(\varepsilon)f(\varepsilon-\mu(H))d\varepsilon,
\end{equation}
so that $C(T,H)=\frac{\partial E}{\partial T}$. Within the
framework of fermionic description, as it is seen from Eq.
\eqref{tll2}, CuPzN is indeed a weakly interacting fermions with
simplest possible spectrum $\varepsilon=p^2/(2m_0)$, where (in
atomic units) $\hbar=$ $c=$ 1. Near QCP taking place at $H=H_s$
and $T=0$, the fermion spectrum becomes almost flat, and the
fermion (spinon) effective mass diverges, $M^*\propto m_0/p_F\to
\infty$, due to kinematic mechanism, for the Fermi momentum
$p_{FH}\to 0$ of becoming empty subband. In case of weak
repulsion between spinons the divergence is associated with the
onset of a topological transition at finite value of $p_{FH}$
signaling that $M^*(T)\propto T^{-1/2}$
\cite{lif,volt,phsc,quasi,khod}. In accordance with Ref.
\cite{rojkov}, we suggest that the weakly interacting Q1DFL in
CuPzN could be thought of as QSL, formed with fermionic spinons,
constituting the Fermi sphere (line) with the Fermi momentum
$p_F$, and carrying spin 1/2 and no charge. For QCP occurs at
$M^*\to \infty$, as we have seen above, we propose that QCP is
FCQPT, at which the corresponding band becomes approximately
flat \cite{quasi,pr,herb2011,book}.

In fermion representation the  ground state energy $E(n)$ can be
viewed as the Landau functional depending on the spinon
distribution function $n_\sigma({\bf p})$, where ${\bf p}$ is
the momentum. Near FCQPT point, the effective mass $M^*$ is
governed by the Landau equation \cite{land,pr,book}
\begin{eqnarray}
\label{HC3} &&\frac{1}{M^*(T,H)}=\frac{1}{M^*(T=0,H=0)}\\&+&
\frac{1}{p_F^2}\sum_{\sigma_1}\int\frac{{\bf p}_F{\bf p_1}}{p_F}
F_{\sigma,\sigma_1}({\bf p_F},{\bf p}_1)\frac{\partial\delta
n_{\sigma_1}({\bf p}_1)} {\partial{p}_1}dv,\nonumber
\end{eqnarray}
where $dv$ is the volume element. Here we have rewritten the
spinon distribution function as $\delta n_{\sigma}({\bf
p})\equiv n_{\sigma}({\bf p},T,B)-n_{\sigma}({\bf p},T=0,B=0)$.
The sole role of the Landau interaction $F({\bf p}_{1},{\bf
p}_{2})=\delta^2E/\delta n({\bf p}_1)\delta n({\bf p}_2)$ is to
bring the system to FCQPT point, where $M^*\to \infty$ at $T=0$,
and the Fermi surface alters its topology so that the effective
mass acquires temperature and field dependences, while the
proportionality of the specific heat $C/T$ and the magnetic
susceptibility $\chi$ to $M^*$ holds: $C/T \sim \chi \sim
M^*(T,H)$ \cite{pr,book,ckz,khodb}. This feature can be used to
separate the solutions of Eq. \eqref{HC3}, corresponding to
specific experimental situation. Namely, the experiment on CuPzN
shows that near QCP at $H=H_s$, the specific heat
$C(T)/T\propto\chi(T)\propto T^{-1/2}$ \, \cite{prl15} which
means that $M^*$ is responsible for the observed behavior, while
QCP is formed by kinematic mechanism. It has been shown that
near FCQPT, $M^*(T)\propto T^{-1/2}$, while the application of
$H$ drives the system to the LFL region with $M^*(H)\propto
(H_s-H)^{-1/2}$ \cite{pr,book,quasi}. At finite $H$ and $T$ near
FCQPT, the solutions of Eq. \eqref{HC3} $M^*(T,H)$ can be well
approximated by a simple universal interpolating function
\cite{pr,quasi,book}. The interpolation occurs between the LFL
($M^* \propto a+bT^2$) and NFL ($M^* \propto T^{-1/2}$) regimes
and represents the universal scaling behavior of $M^*_N(T_N)$
independent of spatial dimension of the considered system
\begin{equation}\label{interp}
M_N^*=\frac{1+c_2}{1+c_1}\frac{1+c_1T_N^2}{1+c_2T_N^{5/2}},
\end{equation}
where $c_1$ and $c_2$ are fitting parameters, $M_N^*=M^*/M^*_M$
and $T_N=T/T_M$ are the normalized effective mass and
temperature respectively. Here,
\begin{equation}\label{MB}
M^*_M\propto (H_s-H)^{-1/2},
\end{equation}
\begin{equation}\label{MT}
T_M\propto(H_s-H),
\end{equation}
are the maximum value of the effective mass $M^*_M$ and
temperature $T_M$, corresponding to the maximum of $(dM/dT)_{\rm
max}(H)$ and/or $\chi_{\rm max}(H)$  \cite{quasi,pr,book}. Below
Eq. \eqref{interp} is used along with Eq. \eqref{tll2} to
describe the experiment in CuPzN.

\begin{figure} [! ht]
\begin{center}
\includegraphics [width=0.47\textwidth]{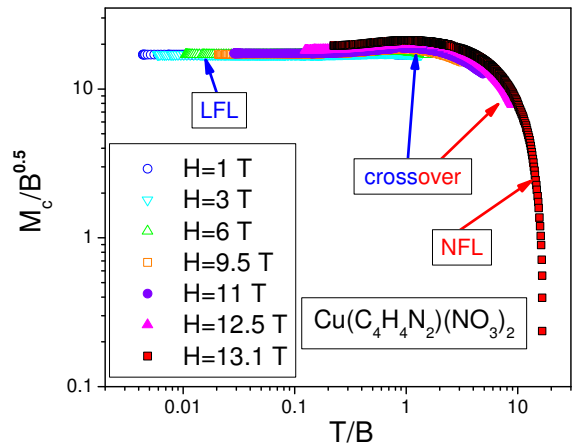}
\end{center}
\caption{(color online). The scaling behavior of the
magnetization $M_c/B^{0.5}$ versus $T/B$ at different magnetic
fields $H$, shown in the legend, with $M_c=a+(M-M_s)$ and
$B=H_s-H$. The LFL, crossover and NFL regions are shown by the
arrows. The data are extracted from measurements
\cite{prl15}.}\label{Fig1}
\end{figure}
Taking into account that $M=\int\chi dH$ and Eqs.
\eqref{interp}, \eqref{MB} and \eqref{MT}, we obtain that
$(M-M_s)/\sqrt{H_s-H}$ as a function of the variable $T/(H_s-H)$
exhibits scaling behavior. This result is in good agreement with
the experimental facts, as it is seen from Fig. \ref{Fig1} that
reports the plot of the scaling behavior of the magnetization
$M_c/B^{0.5}=a+(M-M_s)/(H_s-H)^{0.5}$ as a function of
$T/B=T/(H_s-H)$, with $a$ is a constant added to a better
presentation of the Figure. It is seen from Fig. \ref{Fig1},
that the LFL behavior takes place at $T\ll B$, the crossover at
$T\sim B$, and the NFL one at $T\gg B$, as it is in the case of
HF compounds \cite{pr,book}. It is instructive to note that the
same scaling behavior exhibits $M_c$ obtained in measurements on
$\rm YbAlB_4$ under the application of magnetic field
\cite{s2011}, see Fig. 1(b) of Ref. \cite{jpn2012}.

\begin{figure} [! ht]
\begin{center}
\includegraphics [width=0.47\textwidth]{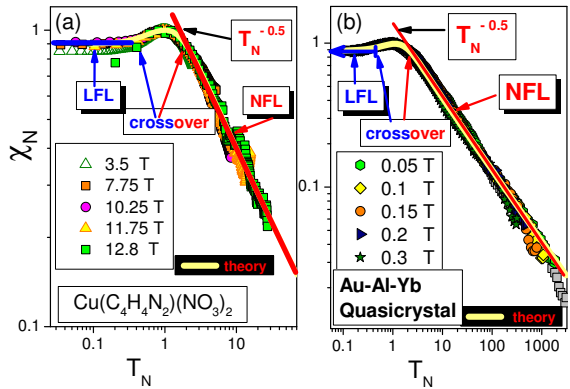}
\vspace*{-0.3cm}
\end{center}
\caption{(color online). Panels (a) and (b). The normalized
magnetic susceptibility $\chi_N$ extracted from measurements in
magnetic fields $H$ (shown in the legend) on CuPzN\,
\cite{prl15} and on $\rm Au_{51}Al_{34}Yb_{15}$ quasicrystal
\cite{yap2}. Our theoretical curves, merged in the scale of the
Figure and plotted on the base of Eqs. \eqref{tll2} and
\eqref{interp}, are reported by the solid lines tracing the
scaling behavior. Panels (a) and (b) show that dependence
$\chi_N(T_N)$ for CuPzN and quasicrystal has three distinctive
regions: LFL, crossover and NFL, where $\chi_N\sim T_N^{-0.5}$
shown by the straight line.} \label{Fig2}
\end{figure}
Figures \ref{Fig2} (a) and (b) portray the comparison between
$\chi_N$ extracted from the experiments on CuPzN, panel (a)
\cite{prl15}, $\rm Au_{51}Al_{34}Yb_{15}$ quasicrystal panel (b)
\cite{yap2}, and the theory. Here $\chi_N$ is the normalized
magnetic susceptibility, while the normalization is done in the
same way as it is done in the case of $M^*_N$ \cite{pr,book}. It
is seen that for more then three decades in normalized
temperature there is very good agreement between the theory and
the experimental data. The double log scale, used in panels (a)
and (b), reveals the universal dependence $\chi_N \sim
T_N^{-0.5}$. The comparison between Fig. \ref{Fig2} (a) and (b)
indicates that $\chi_N$ of both CuPzN and the quasicrystal $\rm
Au_{51}Al_{34}Yb_{15}$ has three regions: low-temperature LFL
part, medium-temperature crossover region where the maximum
occurs, and high-temperature NFL part with the distinctive
temperature dependence $T_N^{-0.5}$. Note that the dependences
from Figs. \ref{Fig2} (a) and (b) qualitatively resemble that of
Q1D HF metal $\rm YbNi_4P_2$ \cite{steg}, and demonstrate the
scaling behavior, and are similar to those of heavy-fermion
compounds \cite{quasi,book}. We recall that the absolute values
of the thermodynamic functions obviously depend on the
interparticle interaction amplitude, therefore, to reveal the
universal properties we have to employ the normalization
procedure \cite{quasi,pr,book}.
\begin{figure} [! ht]
\begin{center}
\includegraphics [width=0.47\textwidth]{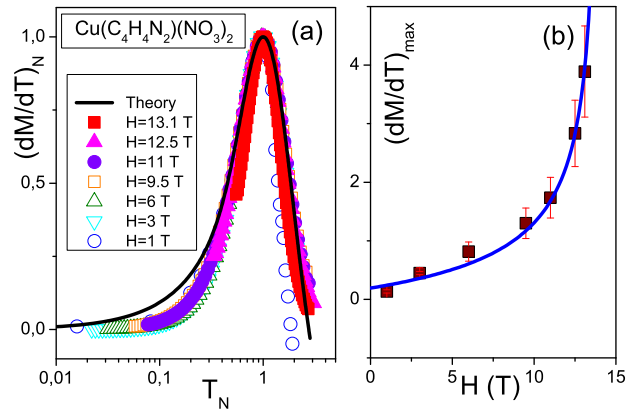}
\end{center}
\caption{(color online). Panel (a): The normalized $(dM/dT)_N$
extracted from measurements in magnetic fields on CuPzN
\cite{prl15}. Theoretical curve, based on Eq. \eqref{tll2} is
also reported. Panel (b) reports the magnetic field dependence
of the maximum values $(dM/dT)_{\rm max}$ of $(dM/dT)$. The
theoretical curve is given by $(dM/dT)_{\rm max}\propto
(H_s-H)^{-1/2}$, see Eq. \eqref{MB}.}\label{Fig3}
\end{figure}
We note, that the approach of weakly interacting Q1DFL
\eqref{tll2} gives for $C/T$ and $\chi$ the same
high-temperature asymptotics $T^{-1/2}$.
\begin{figure} [! ht]
\includegraphics[width=0.47\textwidth]{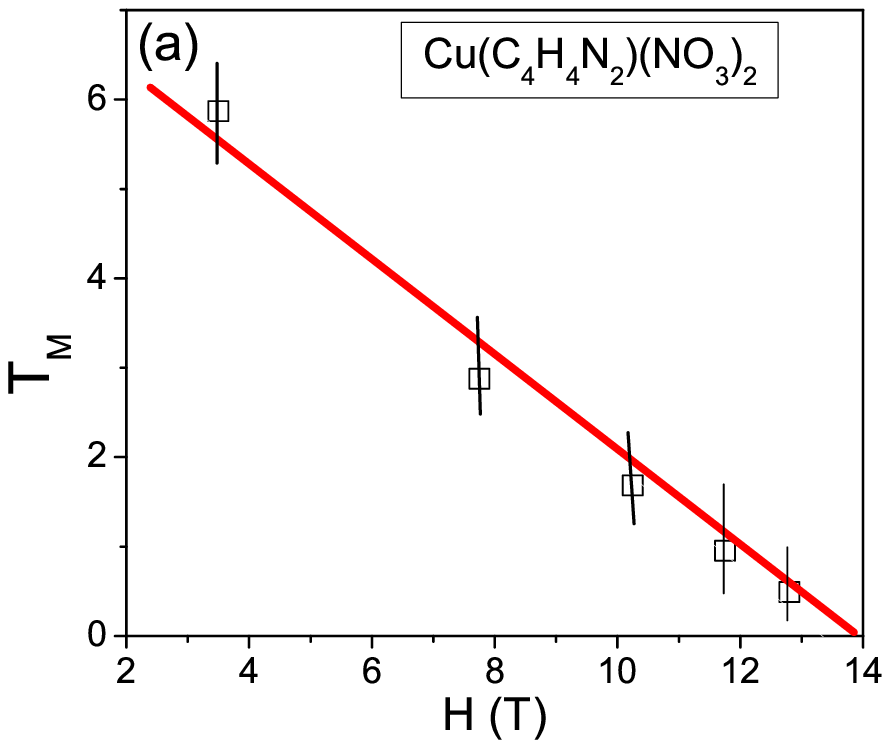}
\includegraphics[width=0.47\textwidth]{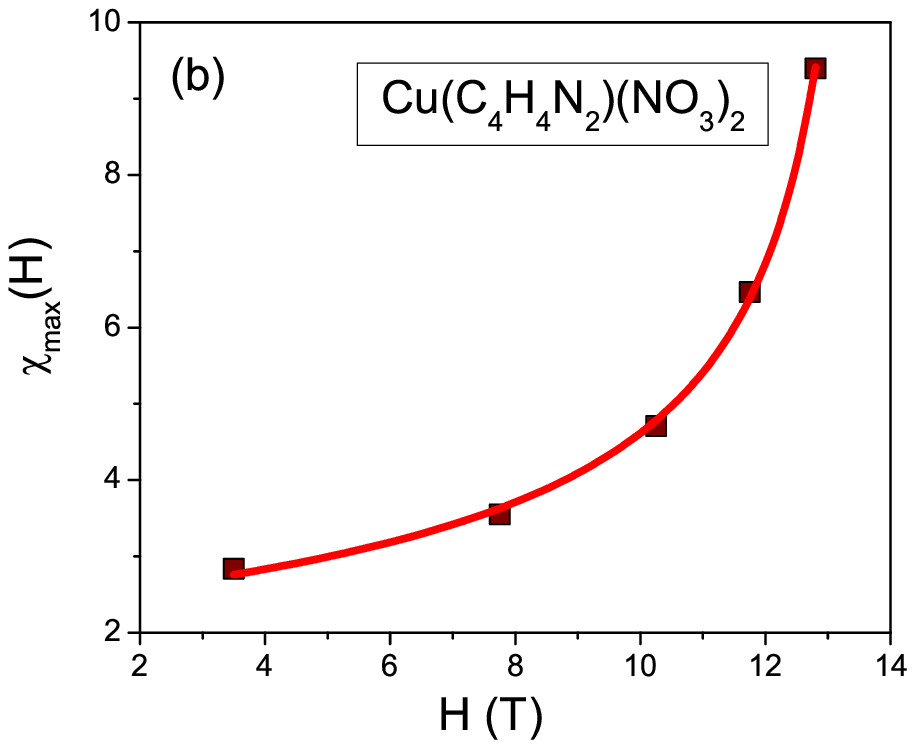}
\caption{(color online). Panel (a) reports the magnetic field
dependence of peak temperature $T_M$ of the magnetic
susceptibility $\chi$ of CuPzN, gathered from experimental data
\cite{prl15}. The calculated straight line $T_M\propto(H_s-H)$
given by Eq. \eqref{MT} demonstrates good agreement with the
experimental data. Panel (b) reports the magnetic field
dependence of the maximum values $\chi_{\rm max}(H)$. The
theoretical curve is given by $\chi_{\rm max}\propto
(H_s-H)^{-1/2}$, see Eq. \eqref{MB}.}\label{Fig4}
\end{figure}

Figure \ref{Fig3} (a) shows the normalized temperature
dependence $(dM/dT)_N$ of the quantity $dM/dT$, revealing the
scaling behavior, while the normalization is done in the same
way as it is done in the case of $M^*_N$ or $\chi_N$. The black
solid theoretical curve corresponds to the temperature
derivative $dM/dT$ of the magnetization \eqref{tll2}. Good
coincidence with experiment on CuPzN is seen everywhere. Such a
good agreement shows that $dM/dT$ has the universal scaling
behavior, that can also be described by taking into account that
$M=\int\chi dH$. In panel (b) of Fig. \ref{Fig3} the maximum
values $(dM/dT)_{\rm max}$ of $(dM/dT)$ versus $H_s-H$ are
displayed. The theoretical curve given by $(dM/dT)_{\rm
max}\propto (H_s-H)^{-1/2}$ is in good agreement with
experimental facts extracted from measurement of the
magnetization \cite{prl15}, and demonstrates that the effective
mass of spinons does diverge at $H \to H_s$.

\begin{figure} [! ht]
\includegraphics[width=0.45\textwidth]{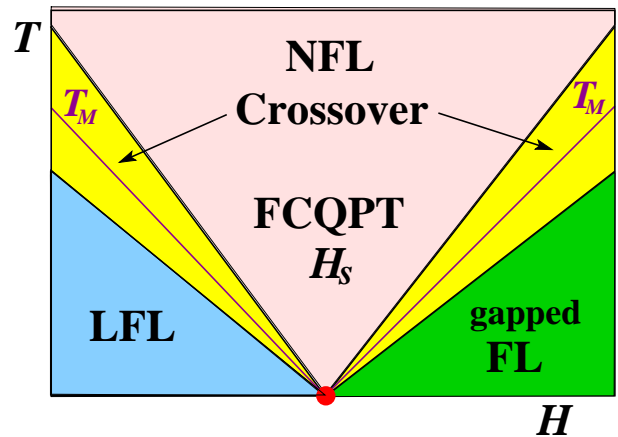}
\caption{(color online). Schematic magnetic field - temperature
phase diagram of CuPzN, based on data from the panels (a) and
(b) of Fig. \ref{Fig2} for $H\lessgtr H_s$. Straight lines on
both sides of $H_s$, which is FCQPT point, indicate,
respectively, the lines of LFL boundary (the lowest
temperature), the temperatures of maxima (middle line, marked
"$T_M$") and the end of crossover region (the highest
temperature at which the system enters the NFL regime), see Fig.
\ref{Fig2} (a). The right sector labeled as "gapped FL" denotes
the gapped field-induced paramagnetic spin liquid. }\label{Fig5}
\end{figure}
The above thermodynamic properties reported in Figs. \ref{Fig2},
\ref{Fig3}, and \ref{Fig4} coincide with those of HF compounds,
and permit us to construct the $T-H$ phase diagram of CuPzN,
shown in Fig. \ref{Fig5}. To do so, in Fig. \ref{Fig4} (a) we
report the peak temperature $T_M$ of magnetic susceptibility as
a function of $H$. It is seen, that the peak temperature $T_M$
goes to zero as $H$ approaches $H_s$. In panel (b) of Fig.
\ref{Fig4} the maximum values $\chi_{\rm max}$ of $\chi$ versus
$H_s-H$ are displayed. It is seen, that theoretical curve given
by Eq. \eqref{MB} is in good agreement with experimental data
extracted from measurement of the magnetization \cite{prl15},
and demonstrates that the effective mass of spinons does diverge
at $H\to H_s$, as it is at FCQPT. The $T-H$ phase diagram
reported in Fig. \ref{Fig5} demonstrates that peak dependence
$T_M$ takes place over wide range of variation of $H$, for
$T_M\propto(H_s-H)$. This shows, that main property of these
lines is that they are straight lines, representing energy
scales typical for HF metals located at their QCP
\cite{scal,sepl}. Since FCQPT takes place at $H=H_s$, the phase
diagram is almost symmetric with respect to the point $H=H_s$,
and consists of the LFL, gapped Fermi liquid, crossover and NFL
regions. The crossover regions in Fig. \ref{Fig5} are shown by
arrows, and are formed by the straight lines, which are the
magnetic field dependencies of temperatures of approximate LFL
and NFL boundaries as well as by that of $T_M$. NFL state occurs
at relatively high temperatures with the distinct temperature
dependence $\sim T_N^{-1/2}$. At the same time LFL regions occur
at low $T$, where the spinon effective mass is almost constant,
as is the case for LFL behavior. At $H>H_s$ the QSL becomes a
gapped field-induced paramagnetic spin liquid, as shown in Fig.
\ref{Fig5}. At rising temperatures and fixed magnetic field $H$,
the system transits through the crossover, and enters the NFL
region, as it is seen from Fig. \ref{Fig5}. It is also seen,
that the crossover becomes wider, as the systems moves from
FCQPT shown by the filled circle. We conclude that CuPzN
exhibits the behavior typical for HF compounds \cite{sepl} that
leads to the formation of the corresponding $T-H$ phase diagram
displayed in Fig. \ref{Fig5}.

In summary, we have shown that the thermodynamic properties of
CuPzN are defined by weakly interacting QSL, and explained the
corresponding experimental facts. Our analysis have shown that
QCP, represented by FCQPT, in CuPzN occurs due to kinematic
mechanism: The band becomes approximately flat not due to
interaction between fermions but rather due to the application
of sufficiently strong magnetic field $H=H_s$. We have
constructed the $T-H$ phase diagram of CuPzN and for the first
time have shown that it is approximately symmetric with respect
to QCP, and has the LFL part, crossover, gapped Fermi liquid,
and NFL part. For the first time, we have also revealed that
CuPzN exhibits the universal scaling behavior typical for HF
compounds.

We thank V. A. Khodel and Y. Kono for valuable comments. VRS is
supported by the Russian Science Foundation, Grant
No.~14-22-00281. K. G. Popov is partly supported by RFBR Grant
No. 14-02-00044 and the Saint Petersburg State University  Grant
No. 11.38.658.2013.

\end{document}